# Pedagogic Challenges in Teaching Cyber Security – a UK Perspective[1]

H.S. Lallie[1*], J.E. Sinclair[1], M.S. Joy[1], H. Janicke[2], B.A. Price[3], R. Howley[2]

[1]University of Warwick, Coventry, CV4 7AL, email: h.s.lallie@warwick.ac.uk

[2]De Montfort University, School of Computer Science and Informatics De Montfort University Leicester LE1 9BH

[3]The Open University, The Faculty of Mathematics, Computing and Technology, Walton Hall, Milton Keynes, Buckinghamshire. MK7 6AA.

[*]Corresponding author

**Abstract**

Cyber security has become an issue of national concern in the UK, USA and many other countries worldwide. Universities have reacted to this by launching numerous cyber security degree programmes. In this paper we explore the structure of these degrees and in particular highlight the challenges faced by academics teaching on them. We explore the issues relating to student expectations and the *CSI effect* in students entering cyber security. We highlight the *science vs tools* debate to bring focus to some of the pedagogic tensions between students/industry and the academics who teach on the degree courses. Cyber security is subject to numerous ethical issues and nowhere is this more so than in a university environment. We analyse some of the ethical teaching related issues in cyber security. This paper will be of interest to professionals in industry as well as academics interested in exploring the shape, flavour and structure of cyber security related degree courses and also the challenges presented to the academics that teach these degrees.

I. INTRODUCTION

Cyber security has been recognised by governments and industry as a crucial factor for ongoing growth, prosperity and safety. In 2010, the UK Government announced that cyber security was a tier 1 priority alongside international terrorism and major accidents [1]. The announcement was supported by a GBP 650 million (USD 1 billion) investment strategy and a promise by the government to "Encourage, support, and develop education at all levels, crucial key skills and R&D", to "Strengthen postgraduate education to expand the pool of experts with in-depth knowledge of cyber [security]" and "Strengthen the UK's academic base by developing a coherent cross-

---

sector research agenda on cyber [security], building on work done by the Government Office for Science." Similarly, the White House had also identified cyber security as a serious economic and national security challenge, and one which the USA did not feel it was currently adequately prepared to counter [2].

In both 2010 and 2011, Barclay Simpson reported high demand for cyber security specialists [3]. By 2011 this demand moved towards candidates with infrastructure and application risk assessment specialisms. Both surveys show demand for penetration testers. In a survey of its members, (ISC)$^2$ (International Information Systems Security Certification Consortium) found that there was strong demand for cyber security professionals and that there was a market belief that job stability and upward mobility is the norm. Despite this, organisations reported that they are "struggling to find qualified candidates" [4].

This has led to an increasing demand for security professionals who have the right skills to tackle ever-increasing challenges. Many universities have responded by developing cyber security degrees at both undergraduate and postgraduate level and some are working with industry to bridge this gap more efficiently.

This inevitably leads to questions as to how best to prepare students: what should be taught, how should it be taught, how can hands-on experience be organised safely and ethically, and how can industry contribute most effectively? Currently there are two approaches taken in the UK Higher Education (HE) sector. Some universities offer dedicated degree courses at both undergraduate and postgraduate level whilst other institutions are providing cyber security content as part of a general computer science degree. This in turn raises curricular questions regarding which part of the cyber security domain should be included.

It appears that there is a strong demand for cyber security graduates and that this is likely to remain strong for a number of years. However the subject as an academic discipline is in its infancy, and this raises further pedagogic and curricular challenges for academics involved in the development and teaching of degree courses. In this paper we begin to explore some of these issues and challenges.

We first consider the current cyber security landscape in UK Higher Education and outline the challenges faced in shaping the curriculum and providing the balance between sound theoretical knowledge and practical skills. These include aspects such as physical resource provision and modes of delivery for distance learning.

We describe the range and "flavours" of degree courses available and explore a number of challenges faced by the academics that develop and teach on these courses. In particular we analyse the often conflicting expectations presented by students and employers and explore the "science" versus tools debate – the two are not mutually incompatible but we highlight the importance of instilling sound scientific principles in students such that they can respond to a rapidly changing landscape. In this discussion we also explore some of the complex ethical issues that are presenting themselves to academics around the world, and in the UK in particular when teaching cyber security. Finally, with learners becoming increasingly mobile, we explore the problems and challenges of teaching cyber security at a distance.

This examination of the pedagogy of cyber security is based on the experiences of the authors who represent a range of UK academic institutions and who lead and teach on a variety of degree courses at both bachelors and masters level. The observations and experiences presented herein are particularly UK centric, however we are confident that many of the issues discussed will resonate with the experiences of academics and professionals in other countries.

The reader should note that the term "course" or "degree course" in the UK means the entire degree programme including all its individual study components (each of which is referred to in the UK as a "module"). In North America and elsewhere the terms "program" and "course" are used to refer to a UK course and module respectively.

## II. THE LANDSCAPE

In September 2011, UK Universities offered 75 security-related undergraduate degree courses showing a substantial development in the provision for Cyber Security/Information Security, Digital Forensics or a combination of both. By September 2012 this increased to 84 degree courses comprising 39 Cyber Security degree courses, 25 Digital Forensics degree courses and 20 Information Security and Digital Forensics degree courses (offered as a combination) - (source: www.ucas.ac.uk/students/coursesearch/). The 39 Cyber Security degrees can be categorised into 20 Computer/Information Security degrees and 19 Network Security degrees, of which the Network Security degrees can be further categorised as 11 Computer Network Security/Network Security degrees, 7 Network Management and Security degrees and 1 Ethical Hacking and Network Security degree. The Computer/Information Security degrees have a range of titles including "Software Development with Data Security", "Mathematics, Cryptography and Network Security", "Computer Security" and "Information Security". There is not much variance in degree names for the digital forensic degrees and "Computer Forensics" or "Forensic Computing" are the most popular names.

In 2012, 56 MSc level degrees in this area were advertised. Of these eight degrees were Digital Forensics related degrees which in many cases contain a number of cyber security modules, four were Information Security and Digital Forensics related degrees, the remaining degrees were Information Security (16), Computer Security (16), Network Security related (10) and Cyber Security related (2) (one of which was "Cyber Security and Management"). Note that in this analysis we are not referring to specific degree titles. Hence the title "Forensic Computing and Security" is loosely categorised as an Information Security and Digital Forensics related degree.

Figure 1 shows specialist degrees which incorporate a large portion of subject specific modules. Typically a cyber security bachelor's degree will incorporate around 4-8 cyber security specific modules over the three years of study. Similarly, half of the modules on a cyber security master's degree will be cyber security specific. The list provided above does not include *generalised* computing degrees such as *BSc Computer Science* or *MSc Information Systems* which may also incorporate cyber security related modules within them. Such degree courses are built around the core aspects of computer science and/or informatics but are designed in such a way to allow for the incorporation of faculty research specialisms. One aspect of UK cyber security education still in its infancy is accreditation. The UK based Forensic Science Society has begun offering two kinds of accreditation for digital forensics, but there are no accreditation systems offered for wider cyber security degrees.

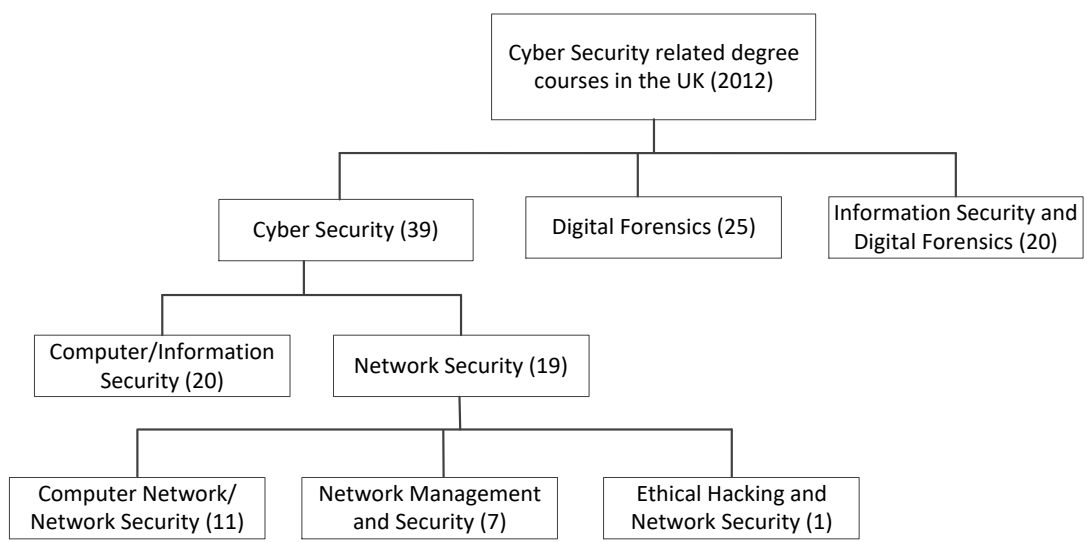

Figure 1. Cyber Security Undergraduate degrees in the UK 2012

III. CURRICULAR CHALLENGES AND STUDENT EXPECTATIONS

A number of the pedagogic problems generally encountered by academics teaching cyber security are often caused by misconceptions and unrealistic expectations of students who enter the study of cyber security at higher education. Cyber security teaching needs to maintain the excitement of the subject and the enthusiasm of these students, while at the same time laying down the academic foundations for in-depth understanding and the development of practical skills.

These unrealistic expectations have been observed in students as follows:

- Students may have an unrealistic perception of the domain and may presume that penetration testing and ethical hacking are the primary and most important fields of cyber security. They fail to understand that strategic issues such as risk analysis, budget control and a deep but well-rounded awareness of organisational culture and context are just as much, if not more important.

- Students may enter with unrealistic expectations of complex cyber security issues such as encryption. There may be a tendency to approach such topics with a *black box* approach, that is data goes in, cipher text comes out. In this case, there is often a resistance to wanting to understand the complexities of protocols and algorithms. There may be little appreciation of the level of mathematical and technical skills required to tackle cyber security effectively.

- Students may have unrealistic views of the careers open to them and misunderstandings of what is actually required in job roles.

Many of these expectations are formed by what is known as the *CSI effect.* The term *CSI effect* was originally used in the area of forensic science to refer to the way that public perception of forensics is influenced by its portrayal in popular television dramas. This is reported to have both positive and negative effects. Public awareness of forensic science has increased over recent years and there has been a growth in the number of students opting to study the subject at university. However, if students enter with unrealistic expectations, the reality of the subject area may lead to problems of disengagement and high drop-out rates [5].

One of the challenges faced in the domain of cyber security is that students may enter without any prior exposure to the subject area resulting in often unrealistic views of what the academic requirements are. They may have exaggerated ideas of what is achievable in the domain and be very surprised by the limitations of science and technology in the cyber security domain. Academics must not only dispel these often strongly held misconceptions but also make students aware of the CSI effect in the colleagues that they may work with in industry. For example, executives and budget holders might expect perfect security for minimal investment driven by a general lack of understanding of the underlying technology used to secure networks.

The CSI effect has also been linked to a belief by degree candidates, and occasionally professionals, in complete certainty and in the infallibility of methods used and results obtainable. Students may enter academia presuming that every problem has an ideal solution and that perfect security is attainable if we get the technology right. One of the important aims of university level teaching of cyber security is to encourage students to question such assumptions and to understand the dangers of binary thinking in which abstract labels of secure/insecure are considered appropriate [6]. Unfortunately, this illusion is encouraged by vendors who promote products as providing "perfect security" or the "ideal security solution".

We can establish from this that the "CSI effect in cyber security" requires particular attention on the part of academics teaching the subject not only to manage and handle perceptions created in students entering the domain, but also to prepare students to handle it in turn when they enter industry. Students will learn that there are many areas with no clear-cut answers, and an abundance of open questions and challenges to be faced. They will discover the need to develop skills and problem solving abilities and to think of security in an holistic way. Far from dampening enthusiasm, this should be seen as opening up exciting intellectual horizons which can lead to challenging and rewarding career opportunities.

IV. THE "SCIENCE" AND ETHICS OF CYBER SECURITY

Given the variety of courses and approaches, the question arises as to what a "good" cyber security syllabus should. In both the specialist and generalised degrees, there are a number of *core curricular concerns* relating to what content the degree course should incorporate. For instance, should cyber security degree students receive a mathematical grounding during their studies? If so, should it extend to a working knowledge of cryptographic algorithms and protocols, or is a *black box* approach sufficient with degree candidates required to understand the fundamentals of cryptography but not the details of the process itself? Should software engineering be a core part of the curriculum? Should degree students undergo a study of computer networks before being introduced to cyber security within the curriculum; or should both be taught concurrently? Or does effective coverage of cyber security issues not require such technical skills?

The answers to some of these questions vary according to context and objective. For example, universities need to decide whether to invest in full undergraduate or postgraduate degree programmes with the aim of sending out future cyber security professionals, or whether to restrict the security offering to one or two modules which are part of a broader computing degree. Discussions between cyber security academics indicate that there are variations in what are considered to be topics and skills required for a meaningful understanding of cyber security. Cyber security perhaps suffers in the same way that computing has in the past - few students coming from high school have studied it and those who have may well have gained an unhelpful impression from the way it is treated in schools. At the other end of the scale, mature students coming from industry to obtain a cyber security qualification may have very specific experience and expectations, and for them the challenge may be in stepping back to examine principles and theories. It may be helpful to move towards a benchmark for cyber security in higher education in the same way that subjects such as computer science are characterised [7].

There are expectations amongst students and employers that a degree course will furnish graduates with a comprehensive ability to use industry standard tools. However, *computer science* academics advocate that the tools will change but the science will not, and therefore students should be given a firm grounding in scientific principles which can then be applied in practice and support continuing professional development (CPD). The argument is similar to one previously applied in software engineering where there was an expectation amongst the student body and employers that students should be able to write programs and use particular programming languages as opposed to an academic view that students should be taught to develop skills which will allow them to analyse and solve complex problems using any programming language.

A good cyber security syllabus must be grounded in firm scientific theory and method. For example, in the area of Digital Forensics, the Daubert ruling requires that for evidence to be admissible in court it must be demonstrably obtained using a scientific method. While this predominantly applies to the US legal system, this will increasingly be the case in the UK where the Forensic Science Regulator is requiring all labs doing forensic work, including digital forensics, to follow rigid scientific quality standards. A consequence is that students entering cyber security must understand and be able to apply in practice scientific principles such as causality. This also means that forensic and security professionals, who may be applying best practices with regard to scientific methods, may not realise that they are applying a formal scientific approach that has its supporters and critics. We support Denning's [8] view that computer science, including cyber security, is a marriage of science, engineering and mathematics and that scientific hypothesis testing is a crucial element of any computing education.

One pedagogic role of universities is to teach students to challenge established patterns and consider the use of alternative and perhaps more suitable tools/techniques and procedures derived from scientific experiments. Deviating from the conventional procedures requires the investigator to demonstrate that the results of their investigation are admissible. This can be done in two ways. Either the validation of the tools and procedures used can be referred to an external body such as NIST. Alternatively the investigator can design, implement and document their own scientific experiments which demonstrate the validity of their approach. Teaching these skills is a fundamental function of universities, in particular at post-graduate level.

Teaching a scientific approach also means that students are equipped with transferable skills for life-long learning and professional practice. Technology is such a fast moving area that the tools students used during their formal education quickly become obsolete, and the students are then required to discover new techniques and learn new tools as part of their professional careers.

This frequently challenges students' perceptions and expectations, formed during traditional schooling, that they should be provided with step by step instructions that can be followed unfailingly. Unfortunately science requires a more nuanced approach and the ability to challenge established conventions in the light of the uncertain world of cyber security. The pedagogic challenge here is to overcome the natural reluctance of students and practitioners to question their established procedures and practices and consider changing them according to the needs of a cyber security issue or incident. Students need mental agility and confidence to deal with the ever complex and unpredictable challenges that will come their way – an education that is purely prescriptive will not adequately support them in meeting these challenges.

Professional bodies publish codes of conduct regulating their members and professions, and which describe what is considered acceptable professional behaviour. In the UK for example, the British Computer Society, the Chartered Institute for IT, in its code of conduct (www.bcs.org/codeofconduct), explicitly refers to "the conduct of the individual, not the nature of the business or ethics of any Relevant Authority". Similarly, there are numerous other relevant codes of conduct such as the "ACM Code of Ethics and Professional Conduct" (http://www.acm.org/about/code-of-ethics) and the IEEE Code of Ethics (http://www.ieee.org/about/corporate/governance/p7-8.html). Module tutors have a challenge in being able to contextualise what are often "general" codes of conduct into the framework and structure of a degree syllabus which focuses on cyber security.

Students on a cyber security degree course may develop personal goals or be presented with technical opportunities that conflict with those of the university – especially where they encounter a module on ethical hacking or penetration testing. Quite often, students develop their own ethical perspective and framework often because the issue of ethics in cyber security (as well as computer science in general) has not been adequately covered early enough in the degree curriculum.

For a number of years, this has raised an ethical paradox for academics who ask whether it is ethical to train a student who may use their skills in an "unethical" manner. Ethics performs an important role in guiding professional practise when novel or unexpected situations are encountered, and one challenge in teaching cyber security is to raise the awareness of ethics amongst the student body.

At the same time, research into cyber security is also shrouded with numerous ethical questions many of which raise similar ethical dilemmas. The anti-circumvention provisions in the 1998 US Digital Millennium Copyright Act and the anti-hacking tool provision in the new Section 3A of the UK Computer Misuse Act [9] have caused concern within the security community, as have proposed EU restrictions on the development of hacking tools. The potential illegality of creating a security penetration tool, to educate students as to how to prevent security breaches, presents us with an interesting ethical conundrum.

Quite often, the problem in defining what is ethically right or wrong within cyber security is as a result of the framework - or lack thereof, within which such interpretations can be placed. For example, the criminality of "Hacking" is often trivialised. Depending on the context in which it is used it may refer to an activity which is, in some sense, potentially "wrong". This may be in a strictly legal sense, such as "unauthorised access" as defined in the UK Computer Misuse Act 1990 [9] and other legislation both in the UK and elsewhere. In general, we can assume that the legal framework in which we are working is underpinned by a commonly agreed ethical framework.

Given the infancy of the subject domain, it is worth considering whether the ethical context of cyber security and digital forensics has developed sufficiently within the profession to an extent that allows us to identify a common framework that guides practice. The answer to this probably is that it has not and that it is time to consider the development of such an ethical framework that builds on and focuses the more general computing ethical frameworks/codes of practice. We believe that there is a need for research into the ethical frameworks that cyber security and digital forensics professionals work within and to better understand how they inform decision making that often poses touch ethical questions.

## V. Teaching Cyber Security at a Distance

The increased profile of cyber security has led to more people seeking a career in this field, many of whom may already be working in a professional environment. For those professionals in this situation, *distance learning* may be a convenient study option.

In the UK, as with much of the world, distance teaching has been much less common than traditional face-to-face teaching. The Open University is of course an exception to this as all its degree offerings have been available through distance learning from the outset. However, in recent years, many universities have started to offer some modules through distance teaching. Despite developments in communication and collaboration tools that have made distance teaching more mainstream, there still remain a number of challenges in the distance teaching of some cyber security subjects.

This mode of delivery introduces particular challenges regarding the use of software and hardware resources which are often expensive, complex and difficult to support and maintain and to make available remotely.

Some of the resourcing problems can be addressed by utilising the growing number of open source applications, for example, to develop penetration testing and ethical hacking experiences. Virtual machine environments may address some of the concerns relating to the potential for network disruption.

For example, in the case of digital forensics, practical experience, in addition to academic knowledge, is required to master many important parts of the discipline. Some of the well-known commercial forensics packages have prohibitive per-seat licences for individuals and currently there is no practical way to share them out at a distance for virtual labs. In this case, open source applications may in fact offer a more comprehensive understanding of fundamental scientific concepts. For instance, rather than becoming a GUI-jockey (e.g. "the tool told me X") students are better able to explain their findings (e.g. "the artefact in this location indicates the user visited the website").

Although many of the practical elements of cyber security education are a challenge to teach at a distance, some of the topics (such as regulation and legislation) are just as well suited to distance teaching. By setting exercises where students debate issues and scenarios in online forums it is possible to acquire a similar experience of the subject as in a traditional classroom setting.

A large part of the intake for distance taught cyber security courses comes from people employed in the IT industry. This means that more advanced exercises can be set using work based scenarios. In the case of topics such as forensic readiness or information security management students can complete coursework based on their real-world experiences.

For intermediate students we advocate having them compile their own tools from source code to create a customised forensic examination machine. This gives them the ability to examine the working of the tool rather than rely on the black box provided by a tool vendor.

In the same way, many other practical topics in cyber security, such as security and penetration testing, can be done safely and inexpensively at a distance using virtualized environments to create local safe copies of entire networks of systems. Network forensics or penetration testing can be done in this virtual world or by connecting the student's examination machine to a remote lab test network via a VPN tunnel.

It is probably true that teaching an entire cyber security curriculum wholly at a distance presents many challenges at the moment. However, a great deal of the introductory and intermediate curriculum is very well suited to modern distance teaching methods and we expect this situation to improve.

### VI. BRIDGING THE GAP BETWEEN ACADEMIA AND INDUSTRY, WHO'S NOT TALKING TO WHOM?

Higher Education policies play an important role in laying the foundation on which future economic and social developments are built. In the UK, the Wilson Review (2012, http://bit.ly/WilsonReview2012) in particular emphasised the importance of the cooperation between industry and academia to equip the next generation with the necessary skills to maintain and advance the economic lead of the country. Other nations have a comparable or an even stronger emphasis on fostering applied research and industry collaboration in university education. The challenge faced by universities is to bridge the gap between laying the foundations of a security-aware culture that understands the wider context in which complex techniques are applied, and the more vocational training on the tools that are used in industry. In highly specialised subject areas such as cyber security and digital forensics, universities are striking a delicate balance between laying a sound knowledge foundation, on which students can build as independent learners throughout their working life, with the training in the latest and often highly complex technologies involved in the sector. The need to provide real-life examples of security incidents, digital forensic challenges and applied research problems often stretches academic capacities, and requires input from industry to support the development of data-sets and realistic examples that are challenging students with problems that are similar to those they will face in their future workplace.

There are a number of collaborations within the UK university sector that are very good examples of how industry and academia collaborate effectively to provide both scientific capacity and professional skills to the student body. An increasingly used pattern of a fruitful collaboration is the development of a new post-graduate academic provision which blends a number of well-established professional certified training courses with a corresponding scientific basis developed as part of a University course. This pattern is particularly interesting for IT professionals that have long experience in the sector and who aim to develop or cross-train in the cyber security area. In our experience this pattern of delivery attracts mainly mature professionals with considerable and successful experience in cyber security and digital forensics, but who do not have a relevant undergraduate degree.

However there are a number of challenges that remain to be addressed in terms of the academic/industrial partnership. The changing higher education landscape in the UK has created pressure for universities to recruit more international students. Most cyber security companies adopt strict recruitment procedures which preclude international students from being able to work as interns, placement or project students due to security vetting requirements. This in turn creates pressures on institutions endeavouring to place their students and in some cases it could be suggested that international students may not receive a similar experience to that of UK/EU students. Nevertheless, an increasing number of international students are overcoming this problem by seeking and securing internships in their home countries via social networks. Universities can positively foster the social networking phenomenon to make it easier for international students to gain internship opportunities both in their home countries and in the EU/USA while continuing to explore ways in which international students can engage more with UK industry.

Another area within the domain of cyber security and digital forensics where industry-engagement is becoming more common is that of projects and placements. Our experience is that industry in this sector is very supportive in developing and providing project opportunities for students. This has a number of advantages for both the industry and partner universities. Industry benefits from a flexible opportunity to transfer knowledge from university developed research into their businesses, without committing expensive resources, as well as being provided with an independent approach to solving their problems. Universities benefit from being able to provide their students with real-life data and problems that complement their academic studies and allow them to relate their knowledge to real industry problems. In this domain, placements and projects frequently involve confidential data which restrict the opportunities, in particular for the international student body.

VII. CONCLUSIONS

Cyber security is a very complex domain and one which is becoming increasingly popular as a career opportunity and as a degree choice. In this article we have highlighted some of the issues faced by UK Higher Education Institutions in providing degree programmes that fill the apparent skills gap. Our survey identifies a wide range of provision characterised by increasing flexibility and support offered to both students and organizations. It presents a picture of steady growth in the number of courses offered and in the range of institutions offering such courses. We also anticipate that ongoing CPD will become a more prominent activity in the learning provision offered by universities in the UK.

We have explored issues relating to curriculum development and delivery that are commonly faced by academics teaching cyber security courses. This is still an area of debate, and one in which stakeholders could usefully come together to provide a subject benchmark. It is inevitable that a new subject area will take time to bed down and that effective curricula and pedagogies will emerge only gradually and with developing experience. However, the subject is now reaching a reasonable level of maturity, and the number of courses on offer may be regarded as a critical mass indicating that a guiding framework would be extremely beneficial. As with computer science, each institution and course is likely to provide its own distinctive approach and "flavour" of provision, but a benchmark ensures some degree of consistency and adherence to a profession-oriented framework. It could also be of use to employers who currently may have little idea of what applicants' degrees may have encompassed.

The need to manage students' expectations and to prepare them in turn to deal with the expectations of others is another area which has to be addressed by continued efforts at education at all levels. Within the UK, it is hoped that awareness of computer science at the school level will increase rapidly due to a variety of factors including government recognition and support, the welcome efforts of organisations such as the grassroots Computing at Schools initiative, and the introduction of good qualifications which assess a strong and interesting computer science curriculum. In the same way, and perhaps building on this new emphasis, a greater appreciation of cyber security issues and what the subject involves would be very beneficial in promoting greater understanding and encouraging school students to consider it as an option for further study and possibly as a career.

In drawing the parallel with the current state of computing in schools it may be worth noting that one of the main difficulties now being encountered is the need to train teachers with the necessary skills to teach the next generation of computer scientists. So with cyber security, good and skilled instructors are needed to deliver the curriculum. This is perhaps another area in which collaboration with industry can be extremely useful, providing cross-fertilisation and practical routes for academics to stay up to date with what is happening in the industry. We would like to see a growing link between universities and the cyber security service industry in the provision of knowledge transfer, placements and mentoring, as well as collaborative research. The challenges of the future are not only found in collaboration and engagement, but also in technology. With the rise of cloud computing many of the traditional approaches to cyber security are challenged and need to be revisited from a much more international stance.

Unconventional modes of delivery, such as distance and flexible learning raise additional and exciting issues relating to the availability of tools and maintaining industrial relevance. While this undoubtedly remains a challenge, we have presented a number of approaches and related them to appropriate areas of the curriculum. Given the growth of distance learning, and indeed open learning, ways to provide effective solutions for supporting non-traditional learners constitute one of the major areas that needs to be developed.

We have noted the difficulties encountered by international students on cyber security courses when placements in the UK are denied for security reasons, and suggest that active engagement with organisations in those students' home countries (perhaps using social networks) can provide internship opportunities for such students. However, this cannot be regarded as a complete solution to the problem and the difficulty caused should not be underestimated, particularly for courses where placement work is an integral part of the degree.

Finally, we note that ethical considerations must underpin any activity involving cyber security (or, indeed, security in general). Students on a security-related course may have their own ethical perspectives which could lead to conflict with permitted behaviour during their studies. It is crucial that, whatever the curriculum, students should be educated in ethical behaviour at the earliest opportunity and that they should be clear about applying their skills in an acceptable and responsible way. Most students have never read their university's acceptable usage policy. Not only must cyber security students be aware of such policies but it may also be necessary for institutions to introduce a further more specifically targeted policy for students on cyber security courses in order to detail very clear expectations for their behaviour both on the course and in their outside activities.

VITAE

Harjinder Singh Lallie (BSc., MSc., MPhil) is a senior teaching fellow in Cybersecurity at the University of Warwick (International Digital Laboratory, WMG). He has previously led courses successfully in Digital Forensics and Security at the University of Derby. His research focus is in the area of Digital Forensics and Information Security and is currently studying towards his PhD.

Jane Sinclair is an Associate Professor in Computer Science at the University of Warwick. She received an MA degree in mathematics and philosophy from the University of Oxford and a PhD degree in computer science from the Open University. She is currently Associate Professor at the University of Warwick. Her research interests focus on formal methods and computer science education.

Mike Joy is an Associate Professor in Computer Science at the University of Warwick. He received the MA degree in mathematics from Cambridge University, the MA degree in postcompulsory education from the University of Warwick, and the PhD degree in computer science from the University of East Anglia. He is currently Associate Professor at the University of Warwick. His research interests focus on educational technology and computer science education.

Helge Janicke gained his PhD from De Montfort University in 2007. Since 2009, Dr. Janicke is working as a Senior Lecturer in Computer Security at De Montfort University. He is leading on subjects relating to access-control, scientific methods and current research topics in security and forensics. Dr. Janicke's research interests are mainly in area of computer security, in particular access control and policy-based system management. Dr. Janicke published his research in many national and international conferences and journals. He is currently co-chairing the Track on Computer Security at the ACM Symposium on Applied Computing (2013).

Blaine Price is a Senior Lecturer in Computing. Blaine obtained his BSc in Computing and Information Science from Queen's University in 1988 and his MSc in Computer Science from the University of Toronto in 1991. He is interested in privacy in mobile and ubiquitous computing, including privacy in location aware applications. He is principal investigator on a number of Knowledge Transfer Partnership projects with industrial partners and co-investigator on PRiMMA (Privacy Rights Managment for Mobile Applications).

Richard Howley is a Principal Lecturer in Digital Forensics and the Course Leader for three post-graduate programmes in the area of forensics and security. After gaining an MSc in Computing Richards interests increasingly focused on the design of secure systems leading to a PhD in this field. Richard has been teaching digital forensics and security related topics for more than ten years as well as supervising several PhD students. His research interests include the procedures and practices of digital evidence acquisition, analysis and management and the professional and moral aspect of the forensic investigations and practices.